\documentclass[preprint,12pt,authoryear]{elsarticle}

\usepackage{amssymb}
\usepackage{color}
\usepackage{textcomp}
\usepackage{gensymb}
\journal{New Astronomy Reviews}

\begin{document}

\begin{frontmatter}



\title{The Case for Jets in Cataclysmic Variables}

\author[1]{Deanne L. Coppejans\corref{cor1}}
\cortext[cor1]{Corresponding author}
\ead{deanne.coppejans@northwestern.edu}
\address[1]{Center for Interdisciplinary Exploration and Research in Astrophysics (CIERA) and Department of Physics and Astronomy, Northwestern University, Evanston, IL
60208, USA}
\author[2]{Christian Knigge}
\address[2]{School of Physics and Astronomy, Southampton University, Highfield, Southampton SO17 1BJ, UK}

\begin{abstract}
For decades cataclysmic variables (CVs) were thought to be one of the few classes of accreting compact objects to not launch jets, and have consequently been used to constrain jet launching models. However, recent theoretical and observational advances indicate that CVs do in fact launch jets. Specifically, it was demonstrated that their accretion-outflow cycle is analogous to that of their higher mass cousins -- the X-ray Binaries (XRBs). Subsequent observations of the CV SS Cygni confirmed this and have consistently shown radio flaring equivalent to that in the XRBs that marks a transient jet. Based on this finding and the emission properties, several studies have concluded that the radio emission is most likely from a transient jet. Observations of other CVs, while not conclusive, are consistent with this interpretation. However, the issue is not yet settled. Later observations have raised a number of questions about this model, as well as about potential alternative radio emission mechanisms. CVs are non-relativistic and many have well-determined distances; these properties would make them ideal candidates with which to address many of our outstanding questions about fundamental jet physics. Here we review the case for jets in CVs, discuss the outstanding questions and issues, and outline the future work necessary to conclusively answer the question of whether CVs launch jets.

\end{abstract}

 
\end{frontmatter}


%%%%%%%%%%%%%%%%%%%%%%%%%%%%%%%%%%%%%%%%%%%%%%%%%%%%%%%%%
%%%%% https://docs.google.com/document/d/11tJPBNjxWe76A-SzHhbIpF5RUinw17TqnGeBbv5gxfQ/edit?ts=5baa56e0
%%%%%%%%%%%%%%%%%%%%%%%%%%%%%%%%%%%%%%%%%%%%%%%%%%%%%%%%%
\section{Introduction}\label{Sec:Intro} 

\subsection{Cast of Characters: Cataclysmic Variables and their Extended Family}\label{sec:cast}

Cataclysmic variables (CVs) are compact binary systems in which a
low-mass main-sequence secondary transfers material to an
accreting white dwarf (AWD) primary via Roche-lobe overflow \citep[for a review see][]{Warner1995}. The
orbital periods of CVs are typically in the range $1.2~{\rm hrs}
\lesssim P_{orb} \lesssim 6~{\rm hrs}$, with a gap in the period
distribution between $\simeq$2~hrs and $\simeq$3~hrs. Stable mass
transfer in CVs, which is characterized by mass ratios $q =
M_2/M_{WD} < 1$, is only possible as a result of angular momentum
losses from the binary. In the standard picture of CV evolution \citep[see][]{Hameury2001,Knigge2011}, 
mass transfer above the period gap is driven by magnetic braking \citep{Eggleton1976,Verbunt1981},
while mass transfer below the gap is driven by gravitational radiation \citep{Faulkner1971,Paczynski1981}.

If the magnetic field of the white dwarf (WD) primary is sufficiently weak ($B_{WD} \lesssim
10^5{\rm G}$), the accretion process takes place via a disk that extends
down to the WD surface. These ``non-magnetic'' systems dominate
the overall CV population and are the WD analogues to low-mass X-ray
binary systems (LMXBs, which contain accreting neutron stars or black
holes). As discussed in Section \ref{sec:XRBs} and in Chapter 7 of this volume, the existence of radio jets in
LMXBs has been well established for quite some time. By contrast,
non-magnetic CVs have only been discovered to be significant radio emitters
-- and possible jet sources -- in the last decade. These systems are
the main focus of this review.

Even though we will not be discussing them in any detail, it is worth noting that there are other classes of AWDs. First, ultracompact  ($P_{orb} \lesssim 1$~hr) AWD binaries harbour
(partially) degenerate donors and are known as  {\em AM~CVn stars}. No radio emission or other jet signatures have so far been reported in these systems -- but we are also not aware of any dedicated searches. Second, AWDs in wide binaries ($P_{orb} \gtrsim 6$~hr) tend to have nuclear-evolved companions, with the longest-period systems ($P_{orb} \sim {\rm years}$) containing red giant secondaries. In these systems with red giant companions, known as {\em symbiotic stars}, mass transfer can be wind-driven. Jets {\em have} been detected in some symbiotic stars, on the basis of radio emission, optical and
X-ray imaging and optical spectroscopy \citep[e.g.][]{Tomov2000,Kellogg2001,Brocksopp2004,Leedjarv2004}. In at least one symbiotic system, radio flares tend to be preceded by 
X-ray dips, similar to what is seen in X-ray binaries \citep{Sokoloski2003}.
Third, AWDs with more massive donors ($M_2 \gtrsim
1~M_{\odot}$) and inverted mass ratios ($q = M_2/M_{WD} > 1$) tend to
display  luminous ($L \simeq 10^{36} - 10^{37}~{\rm erg~s^{-1}}$),
soft ($T \simeq 10^{5}~{\rm K}$) X-ray emission and are therefore 
known as {\em supersoft sources} (SSS). Jets have also been
detected in some SSSs, albeit only indirectly based on blue- and
red-shifted ``satellite lines'' in their optical spectra \citep{Crampton1996,Southwell1996,Tomov1998,Quaintrell1998,Becker1998,Motch1998}.
Fourth and finally, some CVs contain AWDs whose magnetic
fields are strong enough to either truncate the accretion disk
({\em intermediate polars}: $\sim10^5~{\rm G} \lesssim B_{WD} \lesssim
10^{7}~{\rm G}$) or prevent disk formation entirely ({\em polars}: 
$B_{WD} \gtrsim 10^{7}~{\rm G}$). These ``magnetic CVs'' produce
radio emission (likely gyrosynchrotron and/or electron-cyclotron
maser radiation; see section \ref{sec:alternativemechs}), but this is thought to be associated with the magnetized accretion flow, rather than with a
collimated, outflowing jet.

\subsection{Persistent versus Transient Accretion: Nova-Likes and Dwarf Novae}\label{sec:NLsvsDN}

Returning to the non-magnetic systems that are our focus (and hereafter simply referred to as ``CVs''), the long-term light curves suggest
a natural split into two basic sub-classes. The first, the so-called
{\em dwarf novae} (DNe), exhibit dramatic
outbursts, with recurrence times ranging from weeks to decades \citep[e.g.][]{Coppejans2016,Otulakowska-Hypka2016}.
At the onset of a DN eruption, the optical brightness increases
quickly by $\sim$2-8 magnitudes, with a characteristic rise time-scale of
only a day or so. The following plateau phase can then last from days
to weeks. The second class of non-magnetic CVs are known as {\em
nova-likes} (NLs), and these typically remain at roughly constant
brightness.

This split in the non-magnetic CV population is analogous to the split 
in the LMXB population between ``transient'' and ``persistent''
systems. This is no coincidence, as the underlying physics are thought
to be the same. More specifically, the outbursts of both DNe and
transient LMXBs can be understood in the framework of the ``disk
instability model'' (DIM, for a review see \citealt{Osaki1996,Lasota2001}). The DIM \citep{Smak1971,Osaki1974,Hoshi1979,Meyer1981} is based on the observation that accretion disks are
subject to a thermal-viscous instability across a fairly wide range of
mass-transfer rates (bounded above and below by $\dot{M}_{crit,+}$ and
$\dot{M}_{crit,-}$ respectively, \citealt{Smak1984}). This instability is associated with
the sensitive and non-monotonic temperature dependence of the opacity
in partially ionized plasmas (i.e. for $T \simeq 7000$~K). In nova-likes and
persistent LMXBs, the mass-transfer rate from the secondary star is
high enough to avoid the instability, $\dot{M}_2 > \dot{M}_{crit,+}$. 
However, in DNe and transient LMXBs, the mass-supply rate lies in the
unstable range, $\dot{M}_{crit,-} < \dot{M}_2 < \dot{M}_{crit,+}$. As
a result, their disks cycle between a cool,  
largely neutral, low-viscosity state  (``quiescence'') and a hot,  
ionized, high-viscosity state (``outburst''). In the quiescent state,
the mass-transfer rate {\em through} the disk is less than the supply
rate, $\dot{M}_{tr} < \dot{M}_2$. Material therefore piles up in the
disk. This eventually triggers the transition to the outburst state,
in which $\dot{M}_{tr} > \dot{M}_2$.

Strong support for this basic picture comes from comparisons between
the theoretically predicted $\dot{M}_{crit,+}$ and the observationally
inferred long-term average accretion rates, $\overline{\dot{M}}_{acc}
\simeq \dot{M}_2$. These confirm that $\dot{M}_{crit,+}$
neatly separates steady ($\overline{\dot{M}}_{acc}  > \dot{M}_{crit,+}$)
and erupting ($\overline{\dot{M}}_{acc} < \dot{M}_{crit,+}$)
systems, in both DNe and LMXBs
\citep[e.g.][]{Coriat2012,Miller-Jones2013,Dubus2018}. Note that the physics responsible
  for {\em nova} eruptions (a thermonuclear runaway on the AWD) is
  completely different to that driving {\em dwarf nova} eruptions
  (the DIM). Bipolar outflows associated with nova eruptions are common, as is radio emission \citep[e.g.][]{Sokoloski2008,Ribeiro2014}. Since nova outbursts are not accretion-powered, they fall outside the scope of this review and we do not discuss them in detail.   

\subsection{Do Cataclysmic Variables Launch Jets? Motivation and Background}\label{sec:motivationforjets}

Accretion disks are a key component in several astrophysically
important systems on all scales. In young stellar objects, the
accretor is a still-forming proto-star. In CVs, SSSs and symbiotics,  it is a WD. In
X-ray binaries, it is a neutron star (NS) or a stellar-mass black hole (BH). In
active galactic nuclei and quasars, it is a supermassive black hole.

Remarkably, despite these differences in size and mass scales, 
a fair amount of the physics governing the accretion process in all of these objects 
appears to be universal. For example, the statistical properties of
accretion-induced variability (``flickering'') appear to scale
predictably with accretion rate and system size \citep[see][]{Uttley2001,Uttley2005,Gandhi2009,Scaringi2012,Scaringi2015}. Similarly (and paradoxically), the accretion process always
seems to go hand-in-hand with mass {\em loss} from the system, in the form of weakly collimated bipolar disk winds or highly
collimated jets (e.g. \citealt{Greenstein1982, Fender2001,Froning2001,Ponti2012,Kafka2009,Higginbottom2019} and other references in this review).

In an effort to shed light on the mechanism that drives the collimated
jets in these systems, \citep{Livio1997,Livio1999} first drew attention to the fact
that -- at that time -- CVs were the {\em only} class in which no jet
had ever been seen. Weakly collimated disk winds {\em are} observed in
NL CVs and DNe in outburst, but there was no evidence for the
existence of collimated jets in CVs. Based on this apparent failure to launch jets from CVs, \citet{Livio1997,Livio1999} [also see \citealt{Soker2004}]
proposed that jet formation requires a powerful energy source
associated with the central object, which is present in all
disk-accreting systems {\em except} CVs. \citet{Spruit1997}, as well as \citet{Jafari2018}, also noted the absence of evidence for jets in CVs. However,
they interpreted it as a 
consequence of the small dynamic range in radii spanned by CV disks,
since their preferred collimation mechanism (via poloidal magnetic
fields) requires a large dynamic range to achieve strong
collimation. In any case, regardless of any preferred interpretation,
a unique inability of CVs to drive jets would be a powerful constraint
on theoretical models of jet formation and collimation. In particular,
if jets do exist in CVs and display a similar phenomenology to those
in neutron star and black hole XRBs, any universal driving
mechanism requiring strong gravity and/or ultra-strong magnetic fields
would be disfavoured \citep[e.g.][]{Kylafis2012,Parfrey2016}.

In evaluating the detectability of jets in CVs, three observational
signatures are particularly relevant: (i) spatially resolved emission
in optical, radio or X-ray 
images (as seen, for example, in young stellar objects, as well as
some X-ray binaries and AGN); (ii) unresolved radio emission
associated with jet synchrotron radiation (as seen,
for example, in other X-ray binaries and AGN); (iii) jet ``satellite
lines'', i.e. red- and blue-shifted emission line components produced
by line-formation in the approaching and receding parts of the jet,
respectively (as seen, for example, in some SSSs and in the
ultraluminous X-ray binary SS~433). Back in 1997, none of these
signatures had been seen in non-magnetic CVs. Although radio surveys of
non-magnetic CVs in the 1980s yielded five claimed detections [SU~UMa \citep{Benz1983}, but see \citet{Coppejans2016b}, TY Psc \citep{Turner1985}, UZ Boo \citep{Turner1985}, EM~Cyg \citep{Benz1989} and AC~Cnc \citep{Torbett1987}], at least two of these detections were almost
certainly false positives [UZ~Boo \citep{Nelson1988,Benz1989} and AC~Cnc \citep{Koerding2011}] and none were detected in follow-up observations.

Of course, absence of evidence is not (always) evidence of
absence. Indeed, \citet{Livio1997,Livio1999} already emphasized the importance
of establishing the presence or absence of jets in CVs
definitively. Given that optical jet satellite lines {\em had} been seen in SSSs --
another class of AWDs -- \citet{Knigge1998} attempted to estimate
the expected equivalent width of such features in non-magnetic
CVs. Scaling from the jet satellite lines in SSSs, their estimate for luminous
NLs was $EW \simeq 0.1~$\AA. In principle, this is difficult, but
perhaps not impossible to detect in high S/N data.
At radio wavelengths, the main obstacle was simply the
limited instrumental sensitivity. The flux limit for a typical radio
observation at this time was $\simeq 0.2$~mJy, which corresponds to a specific
radio luminosity of $L_{R,\nu} \sim 10^{16} (d/200~{\rm pc})^2$ erg/s/Hz. As we
shall see in Sections \ref{sec:SSCyg} and \ref{sec:caseforjetsdnpop}, this sensitivity turns out to be
insufficient to rule out the presence of radio jets in CVs with
confidence. 

However, in addition to these practical difficulties, the search for jet
signatures in CVs was also constrained by a conceptual issue. As noted 
above, \citet{Livio1997,Livio1999} conjectured that the mechanism responsible
for driving astrophysical jets may {\em require} the existence
of a powerful central energy source. In line with this idea, the
thinking about jets in CVs at the time was focused on the most luminous,
steady CVs, i.e. the highest accretion rate nova-likes. After all,
in these systems, a sufficiently strong and concentrated source of
power might be provided by the hot inner disk and/or a luminous
boundary layer and/or an accretion-heated WD.

This focus -- and Livio's conjecture -- appeared to receive strong
observational support when \citet{Shahbaz1997} announced the
discovery of blue- and red-shifted H$\alpha$ jet satellite
lines in the nova-like T~Pyx. However, T~Pyx is not just any old nova-like,
but also a recurrent nova.
Nova eruptions -- as opposed to the {\em dwarf} 
nova eruptions described above -- occur when the non-degenerate
envelope of an AWD reaches a critical pressure, triggering a
thermonuclear runaway. They are always associated with the
ejection of significant amounts of material, typically $10^{-5} {\rm
M_{\odot} - 10^{-4} M_{\odot}}$, at velocities of $300~{\rm km~s^{-1}}
- 3000~{\rm km~s^{-1}}$. In many recent novae -- including T~Pyx -- this material
produces a detectable, line-emitting ``nova shell'' around the
system. Following up on the
announcement of \citet{Shahbaz1997}, \citet{OBrien1998} and \citet{Margon1998} obtained {\em spatially
resolved}  spectroscopy of T~Pyx and its nebula. These observations
showed that the ``jet lines'' identified by \citeauthor{Shahbaz1997} are, in
fact, nebular [N~{\sc ii}] lines associated with the nova shell. The
only other search for jet satellite lines in high-accretion-rate
NL variables also produced no detections \citep{Hillwig2004}. In the wake of these negative results, progress
effectively ground to a half for several years.

The conceptual leap that rekindled interest in the field
was made by Elmar K\"{o}rding. His approach was driven by the work he
had done on radio jets in black-hole XRBs and AGN \citep{Koerding2007b}. A key
development in {\em that} field was the realization that strong radio
emission is almost exclusively associated with {\em transient}
systems, and is actually quenched in these systems as they transition
towards the peak of their outbursts \citep{Fender2004,Migliari2006,McHardy2006}. Given that the outbursts in
CVs and XRBs are thought to be driven by the same physics (the DIM), he
therefore wondered if these similarities might extend to jet
formation. Could it be that jets were actually present in CVs,
but only (or mainly) in transient systems? If so, the search for jets in
CVs should focus not on the steady, luminous nova-likes, but instead on dwarf
novae away from the plateau phase of their eruptions.

%%%%%%%%%%%%%%%%%%%%%%%%%%%%%%%%%%%%%%%%%%%%%%%%%%%%%%%%%
\section{From Transient XRBs to DNe: The Accretion-Jet Framework}\label{Sec:Simple story}

\subsection{Accretion, Outbursts and Jets in XRBs}\label{sec:XRBs}

Since we will be exploring the similarities and differences between
CVs and XRBs throughout this review, it is worth reminding ourselves
of the ``canonical'' picture of the disk-jet connection that has been
developed for the latter class \citep{Fender2004}. XRBs can be either persistent or transient X-ray sources. As noted
above, this dichotomy can be understood in the framework provided by
the DIM. In this scenario, transient/persistent systems have
mass-transfer rates below/above the critical rate required for steady
accretion \citep[e.g.][]{Coriat2012}. Since both $\dot{M}_2$ and
$\dot{M}_{crit,+}$ depend on the mass of the accretor, it turns out
that XRBs containing BH primaries are almost always transient, whereas
those containing NSs are more equally split between the two
groups. With some exceptions explored further below, radio jets are
mainly observed in transient XRBs, and only in particular phases of
their eruption cycles.

Even though the DIM successfully accounts for the {\em existence} of
transient XRBs, it fails to explain many key features of their
outbursts and spectral transitions \citep[e.g.][]{Coriat2012,Salvesen2013,Nixon2014,Begelman2014}. However, over the last two decades, a clear {\em
empirical} picture has emerged. Briefly, the accretion flow in XRBs
appears to consist of 
two main components: (i) an optically thick, geometrically thin accretion disk
that produces mainly soft X-rays \citep{Shakura1973}; (ii) a hot, optically thin 
``corona'' that produces mainly hard X-ray photons \citep{Shapiro1976} -- the physical nature of the corona remains a mystery (see the Chapter 7 of this volume). Depending on which of these
components is dominant, an XRB can find itself in one of two basic
accretion states: if the corona dominates the X-ray emission, the
system is said to be in a ``hard state''; if the disk dominates, it is
said to be in a ``soft state''.

Erupting XRBs tend to switch
between these states in a consistent and repeatable manner. In
quiescence, most of the luminosity is generated by the corona, while 
the inner edge of the optically thick accretion disk is
located far from the compact object \citep{Esin1997}. The X-ray emission is thus faint and hard. Once the outburst commences, the X-ray luminosity
increases, but the system initially remains in a hard state.
Near the peak of the outburst, the coronal hard
X-rays emission weakens abruptly, while the soft X-rays associated with
the disk brighten rapidly at the same time. This softening may be due 
to the inner edge of the optically thick disk moving inwards, closer
to the compact object \citep{Esin1997}. Overall, the transition from the
corona-dominated hard state to the disk-dominated soft state takes
place at roughly constant bolometric luminosity. The XRB then fades
again, while still remaining in the soft state at first. Eventually,
at a bolometric luminosity of about $0.02~L_{Edd}$ \citep{Maccarone2003},
the system abruptly switches back to the hard state. After this, it
finally fades back into quiescence. Since the hard-to-soft and
soft-to-hard transitions take place at different luminosities,
transient XRBs are said to exhibit {\em hysteresis}. In the
hardness-intensity plane (Hardness Intensity Diagram, HID), the overall outburst
evolution describes a q-shaped trajectory (e.g. the left and middle panel of Figure \ref{fig:Hysteresis_diagram}). For a more detailed description, see Chapter 7 of this volume.

Essentially all observational signatures -- from broad-band
variability, to quasi-periodic oscillations, to the 
presence or absence of disk winds -- depend on the accretion
state of the system \citep[see][for an overview]{Remillard2006}. The radio emission produced by their jets is no
exception \citep[see][]{Fender2004}. In the hard-state, the radio luminosity appears to be
dominated by unresolved synchrotron emission from a compact continuous
jet. As the system starts to rise in X-rays, the radio emission initially
also brightens. During the subsequent hard-to-soft transition near the
peak of the outburst, bright radio flares are observed and are followed by a quenching of the jet core radio emission \citep[e.g.][]{Tananbaum1972,Mirabel1994,Tingay1995}. The phase at which the core jet behaviour changes is often referred to as the ``jet line'' in the literature. As we use this term here, we note the jet line is neither a single line, nor constant: a given system can show this transition in jet behaviour at a range of luminosities and X-ray hardness values. The radio flaring activity appears to be associated with the ballistic ejection of
discrete synchrotron blobs that can sometimes be resolved and tracked
in the aftermath of the state transition \citep[e.g.][]{Mirabel1994,Hjellming1995,Fender1999,Fender2004,Tetarenko2017}. The ejected material
may be associated with the base of the jet and/or the corona; the removal of these regions 
might explain the quenching of the radio and hard X-ray emission. No radio emission from the core is then observed again until the system transitions back to the hard state on the decline from outburst. At this point, the steady, compact jet re-emerges. Note, however, that radio emission from the ejected synchrotron blobs {\em can} persist into and throughout the soft state.

\subsection{From XRBs to CVs}\label{sec:XRBstoCVs}

How can the disk-jet connection in XRBs inform the search for jets in
CVs? Arguably, the key lessons are: (i) compact radio jets are {\em always} seen in 
transient systems; (ii) at least in BH XRBs, they are {\em never} seen in the soft state [\citet{Russell2016} note that in the NS XRBs the radio emission is strongly quenched in the soft state of some systems \citep[e.g.][]{Tudose2009,Miller-Jones2010,Migliari2011,Gusinskaia2017}, 
but not all \citep[e.g.][]{Rutledge1998,Migliari2006}];
(iii) their radio emission is usually strongest during the flaring activity
associated with the hard-to-soft state transition just after the rise
to outburst. More generally, jet line crossings are always associated
with state transitions: steady jets are quenched during hard-to-soft
transitions and re-ignited during soft-to-hard transitions. However,
do CVs even exhibit accretion states analogous to those seen in
XRBs, let alone a jet line associated with switching between these
states? 

Perhaps surprisingly, this is a difficult question to answer. The main
issue is that it is not immediately clear what colours one should
use to construct the DN equivalent of the HID. A similar difficulty
had previously been encountered in attempts to connect the behaviour
of BH XRBs to that of AGN, where the suggested solution was to replace
the ``X-ray colour'' adopted for HIDs in XRBs with a more general
``disk fraction'' parameter that quantifies the relative dominance of
the coronal and disk emission components \citep{Koerding2007b}.

In CVs, the accretion disk radiates primarily in the ultraviolet (UV)
and optical range. Their higher energy emission is usually ascribed to
a boundary layer (BL) between the inner disk and the surface of the WD. In quiescent DNe, this BL is thought to be hot and optically
thin, producing primarily hard X-rays \citep[e.g.][]{Pandel2005,Mukai2017}. In nova-likes and erupting DNe, the BL
is thought to be optically thick, producing mainly extreme ultraviolet
(EUV) and soft X-ray radiation \citep[e.g.][]{Suleimanov2014,Mukai2017}. 

\begin{figure}
    \centering
    \includegraphics[width=\columnwidth]{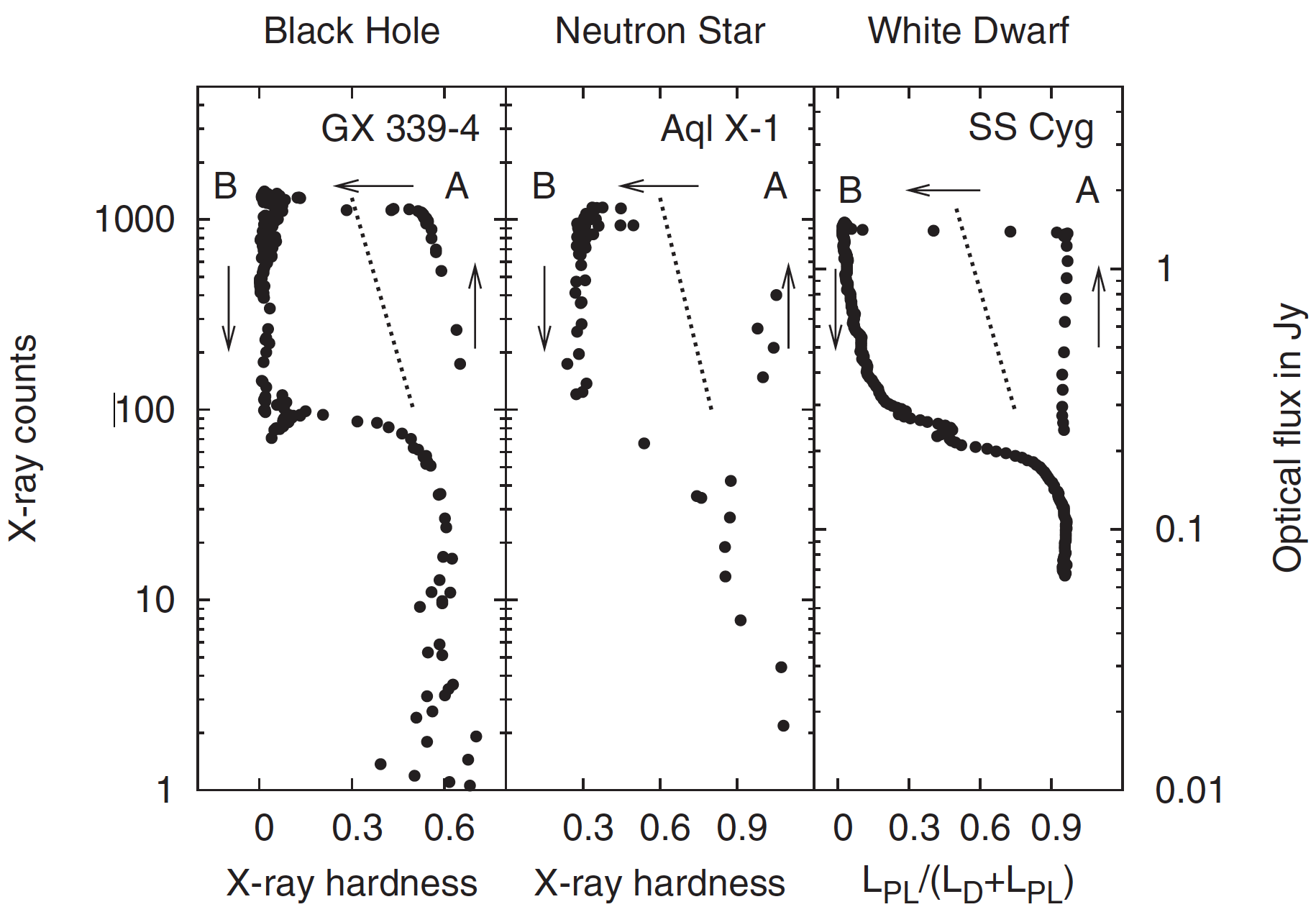}
    \caption{The accretion-outflow cycle of XRBs on a Hardness Intensity Diagram (HID) can be mapped to CVs \citep{Koerding2008}. Specifically, the cycle for SS Cyg is shown on a generalized version of the HID (a Disk Fraction Luminosity Diagram, \citealt{Koerding2006}). In XRBs and CVs the disk emission peaks at X-ray and optical wavelengths respectively; the X-ray (optical) emission is thus a proxy for the accretion power in XRBs (CVs). In CVs the fast moving material in the accretion disk abruptly decelerates and accretes onto the white dwarf via the boundary layer. At low accretion rates the boundary layer is optically thin and emits strongly at X-ray energies, but as the accretion rate increases the boundary layer becomes optically thick and instead peaks in the EUV. The fraction of X-ray emission ($L_{PL}$) emitted by the inner-disk/BL (the sum of the EUV disk/BL emission [$L_D$] and $L_{PL}$) is thus a measure of the optical depth of the inner accretion flow. This ``power-law fraction'' is thus analogous to the X-ray hardness in XRBs which probes the inner accretion flow. Plotting $L_D$ versus the power-law fraction is therefore a generalization of the HID (however, see Section \ref{sec:CV_HID}). By analogy to the XRBs we thus expect a steady jet on the rise to outburst which is quenched (to some degree) and followed by a transient jet when it crosses the jet line (indicated by a dotted line). This transition is marked by radio flaring - like what we observe in SS Cyg \citep{Koerding2008,Miller-Jones2011,Russell2016}. Arrows indicate the progression of the outburst, $L_{PL}$ is the X-ray flux at 3-18 keV, the X-ray hardness in the XRBs is defined as the ratio of the counts at 6.3-10.5 keV and 3.8-6.3 keV and the X-ray counts are taken over the range 3.8-21.2 keV. Credit: Figure 1 from \citet{Koerding2008}. Reprinted with permission from AAAS.}
    \label{fig:Hysteresis_diagram}
\end{figure}

Motivated in part by the observation that (at least some) DNe show an abrupt
switch between hard X-ray and EUV emission during their eruptions \citep{Wheatley2003},
\citet{Koerding2008} considered that {\em this} may correspond to
the state transitions seen in transient XRBs. This would imply that
the optically thin BL in DNe is, in fact, analogous to the
mysterious corona in XRBs and AGN. Based on this idea, they
constructed an HID equivalent for DNe in which ``hardness'' was
measured by the ratio of hard X-rays and EUV luminosities, while
``intensity'' was measured by the optical luminosity. The resulting diagram
for the prototypical DN SS~Cygni is shown in Figure~\ref{fig:Hysteresis_diagram},
alongside the X-ray HIDs for prototypical NS and BH XRB
transients. The similarity between all three diagrams is remarkable. We will return to the physical significance (or otherwise) of this similarity in Section~\ref{sec:CV_HID}. However,
it is worth emphasizing here already that the hysteresis loop for CVs occurs at {\em much} lower Eddington fractions than that 
for XRBs. For example, the bolometric luminosity at the {\em peak} of a DN outburst is typically $\lesssim 0.1\%$ of $L_{Edd}$ \citep[e.g.][]{Smak2000}, whereas the soft-to-hard transition that signals the {\em end} of an outburst in XRBs takes place at $\simeq 2\%$ of $L_{Edd}$ \citep{Maccarone2003}.

Based on the resemblance between these diagrams -- and taking the lessons learnt from XRBs to
heart -- \citet{Koerding2008} concluded that a search for radio jets in
CVs should focus on rising DNe as they cross their putative jet line
near the peak of an outburst. This should provide the best chance of
catching the radio jet at maximum brightness, during the flaring
activity associated with the hard-to-soft state transition. 

The main practical difficulty with this strategy is that the rise
times of DNe outburst tend to be $\lesssim 1$~day. In order to catch
the state transition, it is therefore necessary to monitor a system
optically with high cadence and immediately trigger a sensitive,
fast-response target-of-opportunity radio observation once the onset
of an outburst has been confirmed. This is exactly the strategy that
was executed by \citeauthor{Koerding2008}, based on excellent optical
monitoring provided by amateur observers and coordinated by the American Association of Variable Star Observers
(AAVSO). The very first radio observation
triggered as part of this program led to the clear detection of a radio
flare in the CV SS Cygni (SS Cyg) at the expected phase of its outburst (indicated with red open symbols in  Figure~\ref{fig:Radio_outburst}).

%%%%%%%%%%%
\subsection{SS Cygni: The Prototype}\label{sec:SSCyg}

SS Cyg is a nearby ($114\pm2$ pc) and well-observed dwarf nova. In particular, it is one of the few DNe for which we have detailed UV and x-ray coverage of an outburst \citep[see][]{Wheatley2003}. Using these observations, \citet{Koerding2008} were able to show that the outburst states of XRBs could be mapped to CVs by means of the Disk Fraction Luminosity Diagram (see Figure \ref{fig:Hysteresis_diagram}). To test their prediction of jets in CVs, \citet{Koerding2008} took multiple radio observations over the course of an outburst of SS Cyg.

The observations during the April 2007 outburst confirmed the key predictions for the accretion-jet framework \citep{Koerding2008}. A 1.1 mJy radio flare with a rise time of 1.3 days was detected at 8.6 GHz on the rise the outburst. This subsequently decayed to 0.29 mJy, plateaued for the rest of the outburst, and was not detectable in quiescence. In hard-state XRBs, the radio luminosity and accretion rate follow a relation of the form $L_{rad} \propto \dot{M}_{acc}^{1.4}$, albeit with substantial scatter \citep{Koerding2006}. It is difficult to scale this relation for use with DNe, partly because a ``compactness correction'' of order $M/\rm{radius}$ is presumably required, and partly because it is not obvious which combination of radio flux density and accretion rate should be used to represent the analogue of the bright hard state in a DN. Nevertheless, \citeauthor{Koerding2008} showed that the XRB relation is at least qualitatively consistent with the radio emission seen during the rise and decline of SS~Cyg.

Radio observations of different outbursts of SS Cyg (see Figure \ref{fig:Radio_outburst}) confirmed the radio outburst behaviour \citep{Miller-Jones2011,Miller-Jones2013,Russell2016}, and crucially showed that flare on the rise to outburst is reproducible and consistently detected \citep{Russell2016}. At $\sim$GHz frequencies the initial flare always occurred at approximately 0.5~d after the peak optical brightness, and was typically observed at a few mJy. These observations also showed that the brightness temperature exceeded $10^6$ K, set upper-limits on the polarization fraction of a few percent, and showed lower amplitude variability during the plateau-phase of the outburst. At $5-8$ GHz the spectral index ($\alpha$ where flux density $\propto\nu^{\alpha}$) was steep ($\alpha=-0.77\pm0.14$, \citealt{Miller-Jones2011}) during the initial flare of the April 2010 outburst, and thereafter appeared to flatten over the course of the outburst (although the uncertainty on the spectral index at late times was large - see \citealt{Russell2016}). This was consistent with the flat to slightly-inverted spectral index measured in the April 2007 outburst $0.3\pm0.2$ \citep{Koerding2008}. Higher cadence multi-band observations are needed to strongly constrain the radio spectral evolution. In one VLBI observation taken during the April 2010 outburst, \citet{Russell2016} found marginal ($4\sigma$) evidence for extended radio emission at 0.2-0.5 AU. However, they were unable to conclude whether this extended component was real. The only exception to this general outburst behaviour, was the February 2016 anomalous outburst \citep{Mooley2017}. As the physical mechanism behind anomalous outbursts is unknown and it is not clear whether this outburst can be directly compared to ``normal'' outbursts, we discuss this further in Section \ref{Sec:Simple story}.

Based on the brightness temperature, spectral indices, polarization fractions, variability time-scales and multi-wavelength data, \citet{Koerding2008,Miller-Jones2011,Russell2016} argued that the radio emission in SS Cyg is best explained as synchrotron emission from a transient jet. Alternative mechanisms that produce thermal and coherent emission in CVs have been proposed, but have been ruled out for SS Cyg.

\citet{Cordova1983} and \citet{Fuerst1986} suggested that the wind during outburst could form a cloud of gas surrounding the CV and produce \textit{thermal emission}. There are two arguments ruling out optically thick thermal emission in SS Cyg. i) Optically thick ionized gas typically has a temperature of $10^4-10^5$ K. To produce the observed brightness temperature (the temperature of a black body that would produce the observed radio luminosity), the emitting region would need to be significantly larger than the orbit \citep{Russell2016}. This would be visible at UV/optical wavelengths, but is not seen. ii) More importantly, the observed spectral indices are not consistent with optically thick thermal radiation \citep{Koerding2008}. Optically thin thermal emission, in the form of reprocessed emission or else directly from an outflow, has also been ruled out in SS Cyg. The radio light curve does not follow the bolometric lightcurve, so the former is unlikely \citep{Koerding2008}. Similarly, even if all the material transferred from the secondary star were to be blown off by a wind (clumpy or not) it would only produce $\sim1\mu$Jy - orders of magnitude below the $\sim$mJy detections. For an outflow to produce the observed flux densities (assuming optically thin emission) would require the unlikely scenario where all the transferred material is included in an outflow collimated to an opening angle of $<0.2\degree$ \citep{Koerding2008,Russell2016}.

\textit{Non-thermal emission} in the form of \textit{gyrosynchrotron} or \textit{cyclotron maser} emission produced by magnetically-driven processes such as the disruption of the white dwarf magnetosphere, magnetic reconnections in the disk or reflection of electrons in the converging field lines of the white dwarf have been suggested for non-magnetic CVs \citep{Benz1983,Benz1989,Fuerst1986}. In SS Cyg the flat spectrum and upper-limits on the polarization fraction of a few percent rule out cyclotron maser emission \citep{Koerding2008,Russell2016}. These low polarization fractions also argue against gyrosynchrotron radiation. For gyrosynchrotron emission, there is the additional constraint that in order for the emission to not exceed the compton limit, it must come from an emitting region larger than 10 white dwarf radii. However, the magnetosphere of the WD in SS Cyg is expected to be much smaller than this (less than a few WD radii, \citealt{Warner2002}).  

In contrast, the spectral indices, polarization fractions and high brightness temperatures are all compatible with \textit{synchrotron emission} \citep{Koerding2008,Miller-Jones2011,Russell2016}. This, combined with the temporal and spectral behaviour of the radio emission from SS Cyg led these authors to conclude that the radio emission in SS Cyg is most likely synchrotron emission from a transient jet.

The X-ray emission is another important piece of the puzzle. \citet{Russell2016} found that the radio flare appears to coincide in time with the X-ray flare in SS Cyg. As the optical and radio outbursts of SS Cyg are consistent \citep{Cannizzo1998,Priedhorsky2004,Russell2016}, \citeauthor{Russell2016} showed this by aligning the radio and X-ray observations from separate outbursts based on the simultaneous optical observations. As described above, the X-ray emission in SS Cyg brightens when the increased mass-transfer rate through the disk first drives the material onto the boundary layer. If the radio emission is produced by a transient jet then this association would indicate that the boundary layer plays an important role in jet production in accreting white dwarfs \citep{Russell2016}. Interestingly, during the plateau phase of the outburst, the ratio of radio to X-ray luminosity was such that it could be mistaken for a neutron star XRB -- CVs may therefore masquerade as NS XRBs in surveys that use the X-ray/radio ratio to classify transients \citep{Russell2016}. 

\subsection{The Case for Jets in the General DN Population}\label{sec:caseforjetsdnpop}

Despite these arguments for a jet in SS Cyg, it was not initially clear whether they were applicable to all non-magnetic CVs. Only three non-magnetic CVs (SU UMa, EM Cyg and TY PSc) out of the many that had been observed at radio wavelengths had been detected \citep{Benz1983,Cordova1983,Turner1985,Fuerst1986,Echevarria1987,Nelson1988,Benz1989,Mason2007}, and none of these repeatedly. However, the early phase of the outburst had not been observed at radio wavelengths in any CV prior to SS Cyg and radio observations at this phase were crucial to detect the bright flare. Improvements in telescope response-time, as well as a dedicated optical monitoring campaign with the AAVSO to trigger the observations at the correct time enabled the radio-detection of SS Cyg. Crucially, the Very Large Array (VLA) then underwent an upgrade that significantly increased its sensitivity.

Using the upgraded Karl G. Jansky VLA, \citet{Coppejans2015} and \citet{Coppejans2016} detected a sample of CVs and showed that with modern telescopes, non-magnetic CVs are now a newly-accessible class of radio transient. Typical sensitivity thresholds of previous observations ($\approx0.1-0.2$mJy) were insufficient to detect even the closest CVs (which have flux densities of $\sim$10s of $\mu$Jy). A sample of 5 nova-likes \citep{Coppejans2015}, and 4 DNe in outburst \citep{Coppejans2016} were observed, of which only one nova-like was not detected. At 6-10 GHz the luminosity was in the range $L_{10}\approx4\times10^{15}-4\times10^{16}\,\rm{erg\,s^{-1}\,Hz^{-1}}$ and the emission was highly variable, with measured variability time-scales of $\sim200$s to days. The spectral indices ranged from steep to inverted in the nova-likes, but were unconstrained in the (fainter) DNe. Typical upper-limits on the linear and circular polarization fractions were $\approx10\%$, but one nova-like (TT~Ari) showed a 10-minute flare that was circularly polarized to more than 75\%.  

\begin{figure}
    \centering
    \includegraphics[width=\columnwidth]{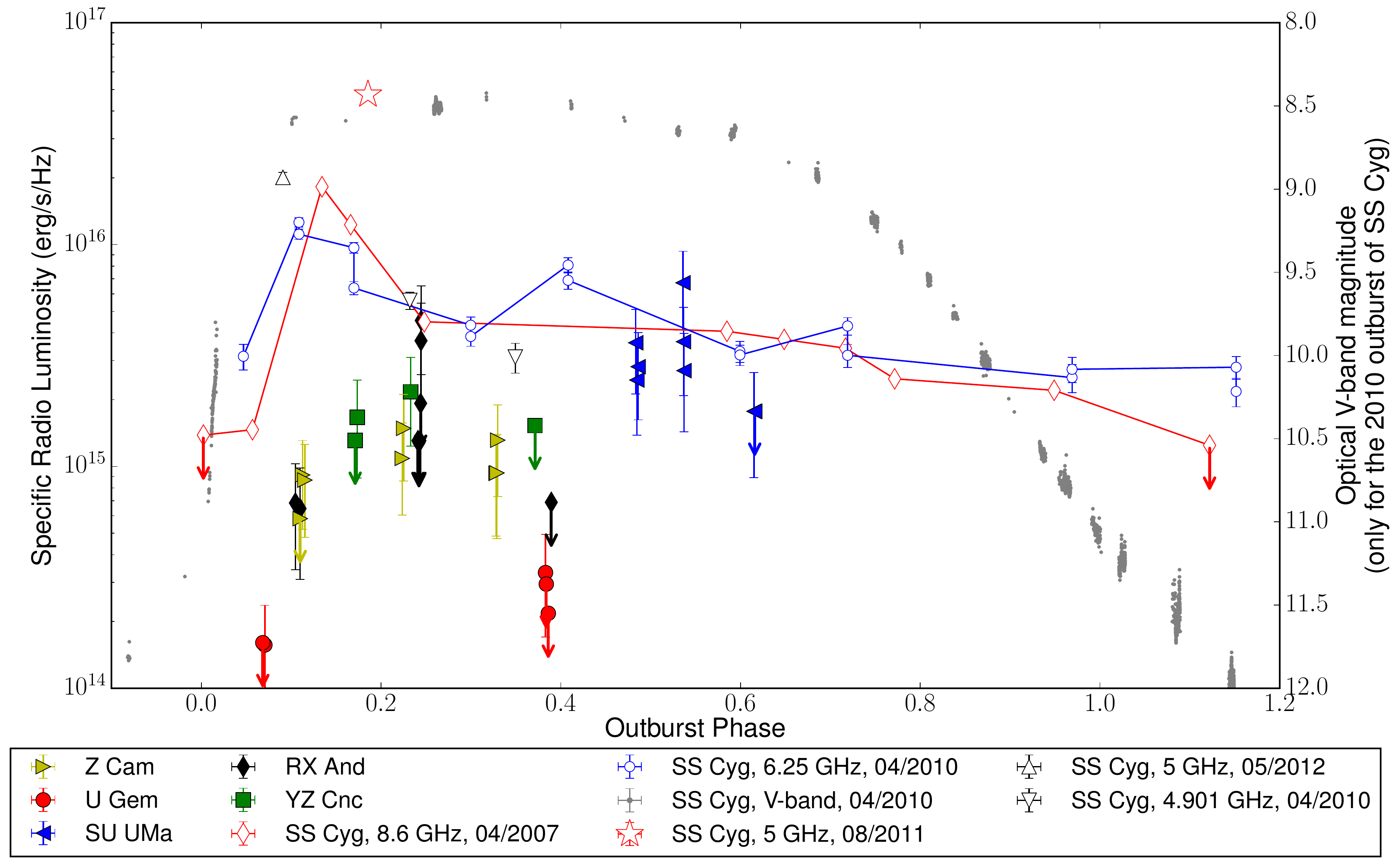}
    \caption{Comparison of the radio and optical light curve behaviour over the course of a dwarf nova outburst. Open symbols and gray points indicate the radio and optical observations of SS Cyg respectively. Filled symbols indicate radio observations of other DNe. As described in the text, SS Cyg consistently shows a bright radio flare on the rise to outburst followed by a plateau for the remainder of the outburst and subsequent non-detections after. This phenomenological behaviour is similar to the XRBs and AGN. In the other DNe, no initial radio flares were detected, although the sampling was too sparse to rule them out. The radio luminosity and variability time-scales of the other DNe were consistent with SS Cyg. Credit: Figure 3 from \citet{Coppejans2016b}, which includes data from \citet{Koerding2008,Miller-Jones2011,Miller-Jones2013} and \citet{Russell2016}. Reprinted with permission from Oxford University Press.}
    \label{fig:Radio_outburst}
\end{figure}

With the exception of TT Ari (which will be discussed further in Section~\ref{sec:alternativemechs}), the radio properties of this sample are consistent with those of SS Cyg and the jet-accretion framework. However, further observations are necessary to conclusively test this and establish the radio emission mechanism. Following similar arguments as outlined above for SS Cyg, the brightness temperature, variability time-scales and polarization fractions indicate that the radio emission in the non-magnetic CVs is \textit{synchrotron emission\footnote{Note that as synchrotron and gyrosynchrotron emission are respectively the relativistic and mildly relativistic form of the same emission mechanism (specifically with reference to the energetics of the electron population), we use the term ``synchrotron'' to refer to both and only distinguish between them when necessary.}} \citep{Coppejans2015,Coppejans2016}. In one of these systems (Z~Cam), mid-infrared emission during outburst has also been interpreted as likely due to synchrotron emission \citep{Harrison2014}. The predicted flare shortly after the rise to outburst is not observed in any of the other CVs (Figure \ref{fig:Radio_outburst}), but given the observing cadence of $\sim1-2$ days, it is not possible to rule it out. As was the case for SS Cyg, the emission was observed well into the plateau-phase of the outburst. The luminosity, variability time-scales and polarization fractions are also consistent with those of SS~Cyg. With the observations to date, the radio emission does not appear to correlate with the optical luminosity, orbital period, outburst type or CV subclass. Given the high variability, higher cadence (and multi-wavelength) observations are necessary to test this however. According to the accretion-jet framework, the radio emission should be dependent on outburst state. In a separate study, Coppejans et al. (in prep.) observed all these DNe in quiescence and made no detections, showing that this is indeed the case. In summary, radio observations of other non-magnetic CVs are consistent with the accretion-jet framework, but further observations are needed to conclusively test this.

\begin{figure}
    \centering
    \includegraphics[width=\columnwidth]{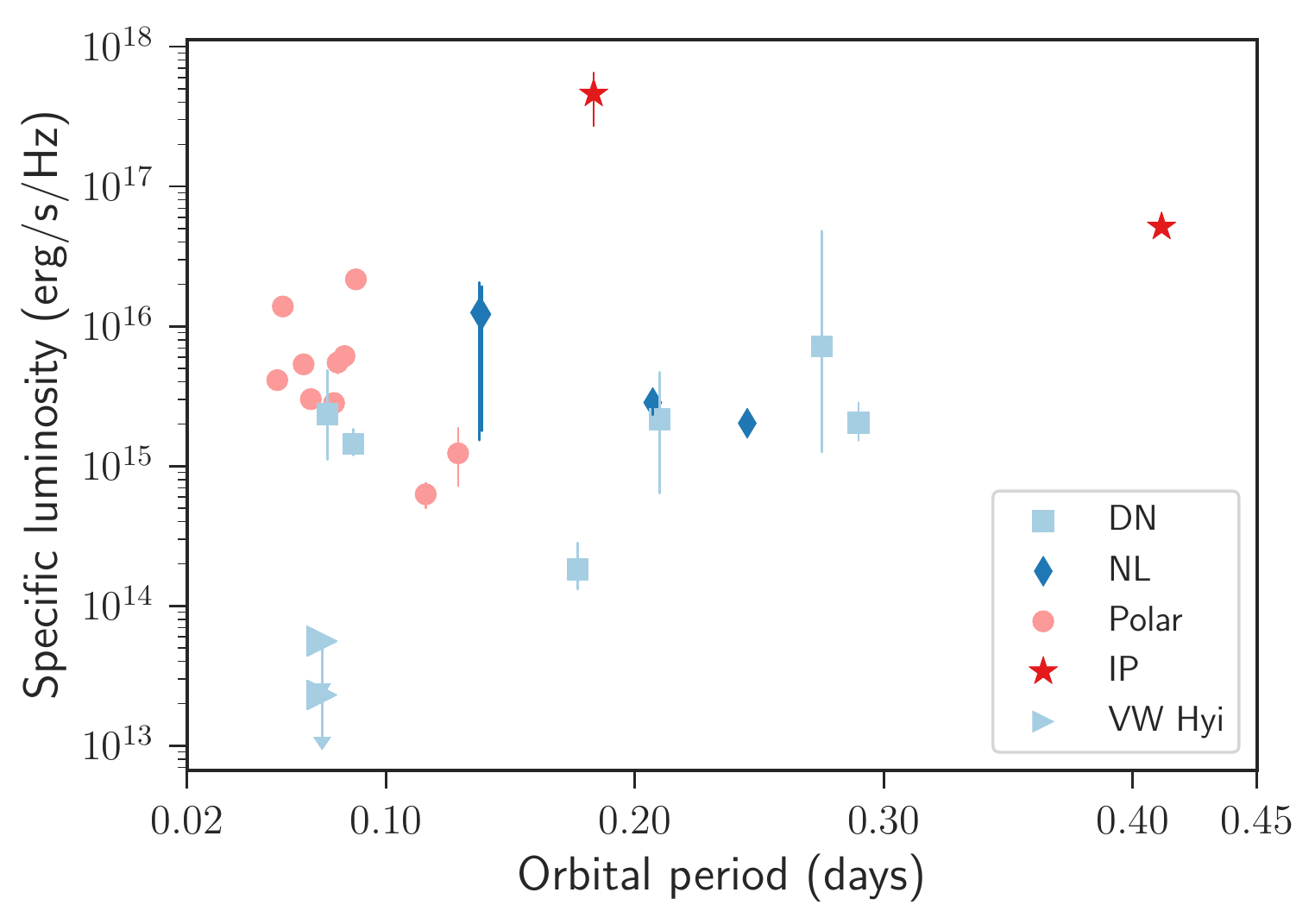}
    \caption{The observed radio luminosities across different CV classes. All recent, sensitive radio observations at $\sim4-12$ GHz of CVs with known distances are included and the bars show the observed variability range of CVs with multiple observations. With current observations, there is no clear distinction between the luminosity of the magnetic and non-magnetic CVs (shown in shades of red and blue respectively). The polar and intermediate polar (IP) classes are defined based on the field strength of the WD: In the former it is $\gtrsim10^7$ G (too high for an accretion disk to form) and in the latter it is $10^{6}\lesssim B_{WD}\lesssim10^7$ and hosts a partial accretion disc. The longest period IP is the unusual system AE Aqr, which shows propeller mode accretion. As the observations of non-magnetic CVs were typically of longer duration and had higher sensitivities than those of the magnetic CVs, a comparison of the upper-limits was not informative and the upper-limits are omitted for clarity. Despite the similar radio luminosities, the two classes do show distinctly different polarization fractions. Polarization in the magnetic CVs is much more prevalent, and at a higher fraction of the total flux density. References: \citet{Coppejans2015,Coppejans2016,Koerding2008,Miller-Jones2011,Russell2016,Barrett2017}; the upper-limits for VW Hyi are from Castro Segura et al. (in prep). The distances used are from \citet{Pala2019}.}
    \label{fig:DN_mag_comparison}
\end{figure}

%%%%%%%%%%%%%%%%%%%%%%%%%%%%%%%%%%%%%%%%%%%%%%%%%%%%%%%%%%%%%%%%%%%%%%%%%%%%%%
\section{The Story is a Little More Complex than that\ldots}\label{Sec:More complex}

\subsection{The Framework so Far}\label{sec:storysofar}

The story we have told up to now is quite logical and coherent. It
begins with the recognition that jets are ubiquitous among 
disk-accreting astrophysical systems. This makes the apparent lack of
jet signatures in CVs interesting and motivates observational efforts
to confirm the presence or absence of jets in these systems. Guided by
theoretical considerations, early searches focus on the
highest-$\dot{M}$ steadily accreting systems, i.e. luminous nova-likes, with
negative results. However, \citet{Koerding2008} then point out
that -- if the disk-jet connection in CVs is analogous to that in XRBs
-- (radio) jets should not be expected in steady nova-likes, but only in the
transient DNe.
Moreover, the brightest radio flares should be seen
near the end of the DN outburst rise phase. They therefore design an
observational campaign to catch such a flare and successfully detect
it in their very first trigger on the proto-typical DN system
SS~Cyg. Finally, follow-up observations confirm that (i) other DNe
also produce radio emission associated with their outbursts \citep{Coppejans2016b}, (ii) no
radio flux is detected in quiescence (Coppejans et al. in prep.), and (iii) SS~Cyg's radio
behaviour is consistent across multiple outbursts \citep{Miller-Jones2011,Russell2016}.

This is extremely satisfying: it brings CVs back into the
fold of other disk-accreting systems, phenomenologically explains their 
radio behaviour, revives the possibility that the disk-jet connection
is {\em universal} and suggests that CVs can be used as ideal
astrophysical laboratories for studying this connection. However, 
it is not the {\em whole} story. As is so often the case, the accumulation of more and
better data has revealed a number of complexities. We discuss these in this section.

\subsection{The Persistence of Radio Emission in the Plateau Phase of DN Outbursts}
\label{quenching}

In XRBs containing accreting BHs, radio emission is quenched in the soft state.
The crossing of the jet line
during the hard-to-soft state transition is associated with bright
radio flaring activity and the ejection of synchrotron blobs. However, 
once this activity has ceased, no persistent core radio emission is seen.
Thus the steady radio jet appears to genuinely switch off in the soft state.

As already noted above, it is non-trivial to identify exact analogues
to ``soft'' and ``hard'' states in DNe (a point we will return to
again in Section~\ref{sec:CV_HID}). However, regardless of where exactly the
transition between these states occurs in DNe, the plateau phase near
outburst maximum clearly belongs to the soft state. In this phase, the
hard X-ray emission is at a minimum, the EUV and soft X-ray emission is
near maximum, and the disk appears to be close to the steady
state expected for a geometrically thin, optically thick accretion 
flow. If DNe behave exactly like BH XRBs, radio
emission should therefore cease during the plateau phase.

This quenching is not observed. As can already be seen even in the
radio light curve from \citet{Koerding2008} (shown in red open symbols in
Figure~\ref{fig:Radio_outburst}), radio emission in SS Cyg persists during the plateau
phase. This behaviour has been studied in more detail by \citet{Russell2016}, whose data show that the residual plateau emission is
typically reduced by about a factor 4 relative to the brightest radio
flares. The detection of radio emission in other erupting DNe \citep{Coppejans2016b} also
suggests incomplete jet quenching. The timing of at least some of
these data sets makes it unlikely that the early bright flare was produced, but the cadence is not sufficient to rule it out.

The reason for the incomplete quenching of radio emission in the plateau phase of DN eruptions is unclear. The persistent radio emission is unlikely to be produced by expanding synchrotron blobs ejected during the earlier flare: it is variable, unresolved and exhibits the usual flat spectral index expected from the compact jet \citep{Russell2016}. Similar behaviour is also seen in some (but not all) NS XRB transients (\citealt{Migliari2004}; also see \citealt{Russell2016}), so perhaps an accretor with a physical surface is required to generate the persistent emission. In any case, this topic is likely to be connected with two other open issues -- the nature of the spectacular flaring activity seen in an anomalous outburst of SS~Cyg, and the origin of radio emission in nova-like variables -- to which we turn in the following sections.

\subsection{Flaring Activity During an Anomalous Outburst of SS Cyg}
\label{sec:lateflare}

There is one radio-observed outburst of SS Cyg that did not follow the mold of the accretion-jet framework: the February 2016 anomalous outburst. Anomalous outbursts are characterised by a long rise time and outburst length \citep[e.g.][]{Cannizzo1998} and show different UV/EUV behaviour during the rise. The standard DIM cannot account for anomalous outbursts, and we do not yet know what the outburst mechanism is \citep[][]{Schreiber2003}. Whether the jet-accretion framework is even applicable in this scenario is therefore an open question. Only one anomalous outburst has been observed at radio wavelengths and the context is not yet clear. In this section we will discuss the unusual radio flaring seen in this event.

The most important deviation from the accretion-jet framework in the February 2016 anomalous outburst of SS Cyg was that no large amplitude radio flare was detected on the rise to outburst \citep{Mooley2017}. Using the flare properties observed by \citet{Russell2016} in a separate outburst, we expect it to have a flux density of $\sim0.8$ mJy at 4.6 GHz and evolve in spectral index from $-0.3\pm0.2$ to $-0.8\pm0.3$. At 15.5 GHz (the frequency of \citeauthor{Mooley2017}'s observations), we would consequently expect a flux density of between 200 and 700 $\mu$Jy. This is consistent with the low-amplitude flaring seen at this phase in the anomalous outburst, so it is possible to explain its absence based purely on the spectral index and observing frequency. However, there are two caveats. First, the initial flare has been detected at up to 5.4 mJy in high time resolution 5 GHz observations \citep{Miller-Jones2013,Russell2016}. This would be clearly detectable at 15 GHz if it was observed at the correct time, especially during the optically thick phase. Second, observations of a subsequent ``normal'' outburst at 15.5 GHz show the predicted bright flare \citep{Fender2019}. It is therefore not yet conclusive whether anomalous outbursts show the predicted radio flare, and indeed whether the disk-accretion framework even applies.

Another peculiarity of the anomalous outburst, was the presence of a bright ($\sim18.0\pm0.4$ mJy) and rapid (15 minute) flare on the decline from outburst phase (compare this to the brightest detection of the initial flare at 5.4 mJy, \citealt{Russell2016}). Towards the end of the outburst the accretion rate drops dramatically, so any scenario attempting to explain this flare with the ejection of material is difficult to reconcile with the large amplitude \citep{Mooley2017}. By aligning this outburst with other SS Cyg observations, \citeauthor{Mooley2017} also found that the flare probably occurred 2 to 3 days before the X-ray emission should peak, so the radio flare is also not likely to be associated with the sudden change in the optical depth of the boundary layer during the return to quiescence. Alternative scenarios have not been explored, but \citeauthor{Mooley2017} suggest that mechanisms like magnetic reconnections or the propeller effect could potentially become dominant at low accretion rates and drive outflows to produce this emission. As the authors point out however, the WD in SS Cyg has a low magnetic field strength. It is thus unlikely that the flare is related to a propeller mechanism and there is no indication of a propeller-like outflow at other wavelengths. No such late-time flare has been observed in the XRBs, but this could potentially be an issue of sensitivity, particularly in the case of the NS XRBs.

The implications of this late flare for the jet-accretion framework are not yet clear. In most of the ``normal'' outbursts of SS Cyg the coverage and cadence was insufficient to rule out a similar flare, but in the April 2016 outburst a similar (although lower amplitude $\approx240\,\mu$Jy) flare was detected \citep{Fender2019}. This flare appeared at a later phase of the outburst however; it appeared after the system had already returned to quiescence, not on the decline. Regardless, large amplitude flaring in the late stages of the outburst may be common \citep{Fender2019} and still needs to be explained in terms of the jet-accretion framework.    
Interestingly, \citeauthor{Fender2019} determined that the energy in the synchrotron emitting plasma of the 18 mJy flare was more than 1\% of the instantaneous radiative luminosity at the time. In contrast, the plasma powering the lower-amplitude flaring throughout the outburst had a time-averaged energy of more than 0.1\% of the instantaneous radiative luminosity. The mechanism behind the synchrotron flaring emission is consequently an important contribution to the kinetic feedback from the accretion process in CVs. Note that this analysis did not make assumptions about the morphology or bulk motion of the outflow and is equally applicable to jets and non-jetted outflows. For comparison, estimates for the jet power in XRBS and AGN are typically more than $\sim10$\% of the instantaneous radiative luminosity \citep[e.g.][]{Maccarone2003b}.

One final point that is worth remarking on, is that rapid flaring was observed throughout the plateau phase of the anomalous outburst. \citeauthor{Mooley2017} measured rise times of between 5 and 30 minutes and flux densities of up to a few hundred micro-Jansky. They state that this behaviour is unlikely to be due to multiple crossings of the jet line, and suggest that it could be due to internal or external shocks, or due to multiple ejection events. Previous observations of ``normal'' outbursts of SS Cyg were taken with a lower-cadence, but also show some flares during the plateau phase \citep[e.g.][]{Russell2016}. Some of the other radio-observed DNe and nova-likes also show highly variable emission \citep{Coppejans2015,Coppejans2016b}. What is producing this variable emission in the high state?

\subsection{Radio Emission from Nova-likes}
\label{sec:nova-likes}

As noted several times already, in transient BH XRBs, radio emission is quenched in the high/soft state. In the hard state, radio emission is consistent with being produced by a compact, steady jet. However, when a source crosses the jet line during its transition from the hard to the soft state, the base of the jet is apparently disrupted and ejected in distinct blobs whose ballistic motion away from the system can sometimes be resolved. These ejections are responsible for the radio flares seen during the hard-to-soft state transition. After the jet line crossing, any residual radio emission is consistent with shocks or these expanding and fading blobs.

This phenomenological picture seems to suggest that radio emission should {\em not} be observed in systems that are persistently in a high/soft state. In CVs, these would be the non-magnetic nova-likes, as the mass-transfer rate from the donors in these systems is above the critical rate that would trigger the disk instability responsible for DN eruptions. Yet, \citet{Koerding2011} and \citet{Coppejans2015} show that nova-likes are, in fact, significant radio emitters, with luminosities that are not too different from those seen in DNe (\citet{Coppejans2016b}; Figure~\ref{fig:DN_mag_comparison}). Is this a problem for a jet-based interpretation of radio emission in CVs?

In a sense, the detection of radio emission from nova-likes should not come as a surprise. As noted in Section~\ref{quenching}, the compact radio emission produced by DNe is actually {\em not} quenched during outburst, and persistent radio flux is {\em also} seen in some (but not all) NS XRB transients. But even among ``steady'' accretors, nova-likes are actually not unique in producing radio (jet?) emission. First, the closest analogues to nova-like CVs in the XRB population may be the persistent Z sources -- neutron stars accreting at a high enough rate to avoid the disk instability \citep{Koerding2011}. Even though they are caught in a permanent high state, Z sources also manage to produce radio emission, usually consistent with an origin in transient, steep spectrum jets \citep[e.g.][]{Penninx1988,Bradshaw1999,Fender2004b,Migliari2006}. Indeed, jet emission has been resolved in the Z sources Cyg X-2 and Sco X-1 \citep{Spencer13,Fomalont2001}. Second, evidence for jets has also been found in bright super-soft sources and symbiotic stars (see the references listed in Section~\ref{sec:cast}). Both types of system contain WDs accreting at a very high rate. This raises the possibility that, at the highest, steady(ish) accretion rates, radio jets may become common again. If so, is the jet-driving mechanism the same as in transient systems?

One surprising aspect of the radio emission seen in nova-likes is the wide range of behaviours they seem to display. For example, their spectral indices vary from flat (or even inverted) to steep, some sources exhibit strong variability, and one source (TT~Ari) displayed $\gtrsim75$\% polarization during a flaring episode \citep{Coppejans2015}. This diversity might point to multiple physical/emission mechanisms being at work (see Section~\ref{sec:alternativemechs}). If so, this complicates the interpretation of the observed radio emission in nova-likes (and also DNe) and makes it even more important to collect more comprehensive data sets for a larger sample of systems.

\subsection{Is the Analogy to XRBs Correct?}
\label{sec:CV_HID}

As discussed in Section~\ref{sec:XRBstoCVs}, the search for radio flares associated with DN eruptions was driven to a large extent by the similarity between the standard HID of transient XRBs and the analogous diagram constructed for SS Cygni \citep{Koerding2008}. To recap: in XRBs, bright radio jet flares tend to occur at a specific location in the HID, during the transition from the hard to the soft state that follows the rise to maximum. If we can construct an analogous diagram for erupting CVs, we might reasonably expect radio flares to occur at a similar location in this. Based on this idea, \citeauthor{Koerding2008} combined optical, EUV and X-ray data to produce the CV version of the HID that is reproduced in Figure~\ref{fig:Hysteresis_diagram}. The similarity between this and the HID for XRBs is obvious, and this directly led to the successful search for a radio flare in SS Cyg, just after the rise to outburst.

But is the CV version of the HID constructed by \citeauthor{Koerding2008} a physically correct analogue to the HID for XRBs? Both diagrams certainly display clear, and phenomenologically similar hysteresis effects -- but are the system components responsible for these effects actually the same? In other words, is Figure~\ref{fig:Hysteresis_diagram} a fair, like-for-like comparison? Does the hysteresis seen in XRBs and CVs have the same (or similar) physical cause?

As pointed out by \citet{Hameury2017}, the answer to these questions is probably ``no''. The fundamental issue is that the diagram constructed by \citeauthor{Koerding2008} for CVs combines a tracer of the inner accretion flow (the X-ray to EUV flux ratio on the x-axis) with a tracer of the outer accretion disk (the optical flux on the y-axis). Hysteretic behaviour in this plane can be understood in the standard DIM framework. Globally, the disk transitions {\em into} the fully ionized state by means of a heating wave and {\em out of} this state by means of a cooling wave. The evolution of the SED produced by the accretion flow therefore depends critically on both the direction (inward vs outward) and the different speeds of these two types of wave front. Different parts of the disk can dominate a given waveband at similar stages on the rise and decline, producing hysteresis. \citeauthor{Hameury2017} show that this is probably sufficient to explain the behaviour of SS Cyg in Figure~\ref{fig:Hysteresis_diagram}. In contrast, both the hard and soft X-rays in XRBs are produced in the {\em inner} accretion flow, close to the central object. This behaviour cannot be explained by the standard DIM -- in fact, the origin of these state changes in XRBs remains one of the most important unsolved problems in the field \citep[e.g.][and references therein]{Salvesen2013}.

Before we move on, we would like to emphasize three points. First, even if any hysteresis seen in DNe turns out to be fundamentally different from that seen in XRB transients, the radio emission in both types of systems may still be produced by jets. However, it would certainly weaken the empirical case for jets in CVs, making the search for {\em resolved} jet emission even more important in these systems. Second, our poor understanding of the corona in XRBs, and the boundary layer in CVs represents a major obstacle for any attempt to compare these types of systems. Is the corona really a corona, or is it an advection-dominated accretion flow (ADAF) or the base of the jet? Or is it some combination of these? Is the optically thin BL associated with the X-rays produced by quiescent DNe really physically different to the corona in XRBs? Or are we just using different names for the same things? Third and finally, \citet{Hameury2017} have presented an alternative HID-analogue for DNe, constructed solely from X-ray and EUV fluxes. These are both associated exclusively with the inner accretion flow -- albeit the poorly understood BL -- yet still appear to display hysteresis.

Whether, or to what extent, the X-ray, EUV and optical outburst behaviour of SS Cyg is proto-typical is currently unclear. For example, some DNe do not seem to exhibit any hard X-ray supression in outburst \citep[e.g.][]{Neustroev2018}, nor is there evidence for a hard X-ray flare during the rise and/or decline in at least some systems \citep{Fertig2011}. However, SS Cyg is unfortunately still the only DN for which simultaneous, time-resolved X-ray, EUV and optical observations that cover a full outburst. Obtaining additional data sets of this kind should be a high priority for the field.

\subsection{The Non-detection of VW Hyi}\label{sec:VWHyi}

There is one DN that does not appear to be a radio emitter in outburst, in stark contrast to the predictions of the accretion-jet framework. Deep radio observations of VW Hydri (VW Hyi) during outburst and superoutburst have yielded non-detections that are one to two orders of magnitude fainter than other DNe in outburst (see Figure \ref{fig:DN_mag_comparison}). VW Hyi is otherwise an archetypal DN, showing outbursts approximately every 25-30 days and superoutbursts every $\sim180$ days \citep{Otulakowska-Hypka2016}. Importantly, at only 54 pc \citep{Pala2019} it is the closest known DN in the Southern Hemisphere. It was therefore surprising when ATCA (Australia Telescope Compact Array) observations at 5.5 and 9 GHz during a superoutburst, and MeerKAT observations (at 1.4 GHz) during a normal outburst yielded non-detections (Castro Segura et al. in prep).

At present it is unclear what the implications for the jet-accretion framework are. The radio observations during superoutburst were taken prior to the optical peak (earlier than the predicted flare), but the observations during outburst were taken shortly after the rise. Based on the cadence of the observations it is still not possible to rule out a radio flare. However, the deep upper-limit on the ``plateau'' emission indicates that DNe show a wide range of radio luminosities. A larger sample of radio-observed CVs are necessary to place this non-detection in context.

\subsection{Are we Seeing Jets at all? Alternative Radio Emission  Mechanisms}\label{sec:alternativemechs}

The previous sections focused exclusively on the jet-accretion framework. However, there are other mechanisms that could potentially produce radio emission in non-magnetic CVs. We will now discuss these alternative mechanisms. As a note to the reader, much of the work (and the proposed mechanisms) that we will refer to in this section were conducted (or proposed) prior to the new high-sensitivity observations showing that non-magnetic CVs are radio emitters, and also prior to the significant expansion of the sample of radio-detected magnetic CVs \citep{Barrett2017}. A number of points we make here are therefore speculative, and much of this section is spent outlining questions and new observations and theoretical developments that are needed in the upcoming years to test these theories.

The obvious place to start is the detection of coherent emission from the nova-like TT Ari \citep{Coppejans2015} that cannot be associated with a jet. Specifically, a $\sim10$ minute flare that was more than 75\% circularly polarized was observed. This flare is not unique among non-magnetic CVs either, as an earlier lower-sensitivity observation of EM Cyg also showed a highly circularly polarized ($\sim81\%$) flare \citep{Benz1989}. Although this does not preclude a jet entirely, it does complicate the story and either calls for the operation of multiple radio emission mechanisms and/or a scenario where a subset of CVs do not launch jets. As some NS XRBs appear to not launch jets (e.g. \citealt{Fender2000}, but also see \citealt{vandenEijnden2018}) this latter claim is plausible, but we do not have a sufficiently large sample to speculate further. The (short) observations of TT Ari do, however, hint at the presence of two types of emission: highly circularly polarized flaring emission superimposed on a slowly varying component with a low fractional polarization. This can be seen in an observation of TT Ari taken 19 hours after the flare, where the emission varied slowly from $250-200\,\mu$Jy over 40 minutes and had a polarization fraction of $\sim10$\% (and is consistent with synchrotron emission). Interestingly, the non-magnetic CV V603 Aql also shows similar behaviour, with varying levels of circular polarization from $<13\%$ \citep{Coppejans2015} up to $69\pm21$\% at 8 GHz \citep{Barrett2017}. This duality of the radio emission properties is very reminiscent of that seen in the magnetic CVs.

In the magnetic CVs, highly circularly polarized flaring emission and slower-varying emission with low levels of circular polarization are seen, and individual CVs can switch between these states and even go through periods where they are not detected at all \citep[e.g.][and references therein]{Chanmugam1982,Dulk1983,Bastian1985,Mason2007,Barrett2017}. In order to not confuse the terminology with that used for the outbursts of non-magnetic CVs, we will refer to the flaring emission as ``bursting'' and the slowly-varying emission as ``quiet-state'' emission in this review (for comparison purposes note that the terminology used in the literature is ``outburst'' and ``quiescent''). When discussing the source of the radio emission in the magnetic CVs, there are two components to consider: the radio emission mechanism itself, and the mechanism that heats and accelerates the electrons. For the quiet-state emission, the proposed mechanisms include gyrosynchrotron emission from electrons trapped in the magnetosphere of the WD \citep{Dulk1983,Chanmugam1982,Chanmugam1987} or the red dwarf secondary star \citep{Dulk1983,Chanmugam1987}. Bursting emission is generally proposed to be electron cyclotron maser emission \citep{Melrose1982} operating near the surface of the WD or the red dwarf secondary \citep[e.g.][]{Dulk1983}. \citet{Kurbatov2018} also put forward a model invoking cyclotron radiation from electrons in a fluctuating magnetic field (driven by Alfv\'{e}n wave turbulence) in the accretion stream. These mechanisms all require energetic electrons. These have been suggested to be produced in the interaction regions of the magnetic fields between the two stars \citep{Dulk1983, Chanmugam1987, Mason2007, Gawronski2018}, through magnetic reconnections near the surface of the red dwarf secondary star [stellar flares] \citep{Dulk1983,Mason2007}, or in a unipolar inductor model in the magnetosphere of the secondary star (\citealt{Chanmugam1982}, but see \citealt{Mason2007}). For completeness, the unusual magnetic CV AE Aqr shows propeller state accretion where radio synchrotron emission is produced in expanding magnetic blobs of material that are expelled from the system by a rapidly rotating WD magnetosphere \citep{Meintjes2005}.

Historically, the radio emission mechanisms for the magnetic and non-magnetic CVs were treated differently, as indeed we have done in this review. However, given that non-magnetic CVs have now also been established as radio emitters and three of them to date show radio properties characteristic of the magnetic CVs, is this justified? It is somewhat tempting to argue that TT Ari, EM Cyg and V603 Aql are just misclassified as non-magnetic CVs. Below $\sim10^6$ G the magnetic field strengths of accreting WDs cannot be directly measured with current techniques. The one possible exception to this statement is that CVs with a very specific combination of accretion rate, WD spin period and WD magnetic field strength could drive magnetically gated accretion which would allow for field strength determinations \citep{Scaringi2017}. TT Ari, EM Cyg and V603 Aql do not have these properties, and importantly do not show any signs of magnetically channelled accretion such as coherent optical/X-ray spin pulsations, Zeeman splitting of the absorption lines in optical spectra, or polarized IR/optical cyclotron emission \citep[see][]{Benz1989,Coppejans2015}. For completeness we do note that there is potentially a level where the magnetic field of the WD is high enough to have a small dynamical impact on the accretion flow and not be detected with current techniques, but given the difficulties in testing this and the fact that about a third of the radio-detected non-magnetic CVs would need to have magnetic fields in this specific range if this is the source of the radio emission, we do not pursue this idea further here. 

At present, there is no clear distinction between the radio luminosities of the magnetic and non-magnetic CVs (see Figure \ref{fig:DN_mag_comparison}). The two IPs appear to be brighter, but one is the highly unusual AE Aqr. The other (V1323 Her) is a factor $\sim100$ times brighter than the average CV. The faintest non-magnetic system, VW~Hyi, is more than a factor 100 times fainter than the average. However, it is too early to know whether this tells us something important, or whether it merely indicates that the luminosity range of CVs is large. A cursory look at the polarization properties of recent high-sensitivity observations show that (at 8 GHz) circular polarization was detected in approximately $\sim30$\% of non-magnetic CVs \citep{Benz1989, Coppejans2015, Coppejans2016b, Barrett2017} and $\sim80$\% of magnetic CVs \citep{Barrett2017}. There are undoubtably important selection effects in these samples (for example the integration times, observing cadence and sensitivity differ markedly between non-magnetic and magnetic CVs and will greatly affect the chance of catching flaring emission), but this seems to indicate that the polarization properties of the two populations differ. This needs to be confirmed in a larger sample of systems, with the selection effects properly accounted for.

If we were to assume that the radio emission mechanisms were the same in the magnetic and non-magnetic CVs, then those requiring large WD magnetic fields would be eliminated. In this case, magnetic reconnections near the surface of a (flaring) secondary star producing cyclotron maser emission and gyrosynchrotron emission from these trapped electrons is a promising mechanism. There is indeed indirect evidence for magnetically active secondary stars \citep[e.g.][]{Richman1994,Baptista2003,Kafka2008}. This would be interesting from at least two fronts. First, magnetic braking via the ionized wind of the secondary star is an important angular momentum loss mechanism in CVs \citep{Eggleton1976,Verbunt1981}. The evolutionary model of CVs is based strongly on the idea that the secondary loses mass over time and eventually reaches a mass where it becomes convective; at this point there is an abrupt drop (or cessation) of magnetic braking \citep{Rappaport1982,Spruit1983} due to the associated change in the magnetic field \citep[see][and references therein]{Taam1989,Warner1995,Knigge2011,Garraffo2018} and subsequent angular momentum loss is driven primarily through gravitational radiation \citep{Faulkner1971,Paczynski1981}. If radio emission probes the magnetic field of the secondary star, then radio observations would be an important tool to probe this physics. Second, among the M dwarfs are a large number of systems (isolated or in binaries) that show bright multi-wavelength flares from reconnection events near their surfaces \citep[e.g.][and references therein]{Osten2005,Fender2015,Schmidt2019}. If the M dwarfs in CVs (in TT Ari the secondary is an M3.5 dwarf, \citealt{Gaensicke1999}) are flaring like these ``flare stars'' then they offer a regime in which to study the dynamo mechanism and magnetic structure in low-mass dwarfs at otherwise inaccessible rotation rates. Rotation rate plays an important role in the stellar dynamo (e.g. see \citealt{McLean2012} and references therein), and the secondary stars of CVs are tidally locked to the spin period of the binary and hence have a spin period of only a few hours. 

\citet{Coppejans2015} looked into the idea of a flaring secondary star in TT Ari and other non-magnetic CVs. The radio luminosity of most of the general (non-flaring) population of non-magnetic CVs exceeds the upper-edge of the quiescent flare star population ($10^{14}$ erg/s/Hz; \citealt{Gudel1993}) by more than an order of magnitude (see Figure~\ref{fig:DN_mag_comparison}). The flare in TT Ari peaked at $\approx38$ times the brightest (typical) M-dwarf flares \citep{McLean2012} and was approximately 10 times more luminous than the super flare of the M dwarf EV Lac \citep{Osten2005}. We do note however, that it had a comparable luminosity to the brightest-ever flare detected from an M dwarf binary - DG CVn \citep{Fender2014}. The radio emission for M0-M6 type dwarfs saturates \citep{McLean2012} at rotation rates of vsini$\approx$5 km/s. For comparison, a rough estimate for a CV secondary star with a 3.3 h orbital period and a radius of 0.4 solar radii is approximately 160 km/s at the stellar radius. Based on this, \citeauthor{Coppejans2015} concluded that the radio emission in TT Ari is unlikely to be due to flares in the secondary star. However, can we directly compare the magnetic fields and reconnection events in stars that have spin periods that differ by a factor $\sim30$, particularly given that the dynamo mechanism in this spin regime is not well understood? There are stronger arguments against this scenario based on the timing of the radio emission however. First, no non-magnetic CV has been detected outside of outburst. Even dedicated follow-up observations of all the radio detected DNe yielded no detections in quiescence (Coppejans et al. in prep.). As described in Section~\ref{Sec:Intro}, outbursts are driven by an instability in the accretion disk. Why then, would we expect the radio emission to be dependent on the outburst state if it is associated with the secondary star? Even if irradiation of the secondary star is taken into account, then how can the large amplitude flare at the start of the outburst in SS Cyg be explained? Would the irradiation have such a dramatic effect on the magnetic field of the secondary star? Second, if a flaring secondary star is responsible for the emission, then we should not see a marked difference in the radio emission properties between the non-magnetic and magnetic CVs (also see \citealt{Mason1996}). All these questions are speculative and need to be revisited with detailed modelling in light of recent detections.

Two other possible sources of radio emission have also been suggested in CVs. The first is cyclotron maser emission from thermal electrons in a fluctuating magnetic field (driven by Alfv\'{e}n wave turbulence) in the accretion stream \citep{Kurbatov2018}. The authors have tested this scenario on the polar AM Her, but state that there is no fundamental reason why this mechanism should not work in non-magnetic CVs. The second mechanism is magnetic reconnections in the accretion disk. \citet{Meintjes2016a,Meintjes2016b} have proposed that the radio emission in CVs could be produced by reconnections that accelerate electrons and produce synchrotron emission via the van der Laan process. Indeed, magnetic fields are known to play an important role in the accretion discs of CVs, as the magnetorotational instability \citep{Balbus1991} drives turbulence. The dependence of the radio emission on outburst state can also qualitatively be explained with this mechanism, as the reconnection rate should depend on the mass transfer rate through the disc. If the radio emission in CVs is produced by magnetic reconnections, then their proximity would make them promising laboratories to study magnetic reconnections and other magnetohydrodynamical processes in accreting systems \citep[e.g][]{Meintjes2016b}. Further simulations and quantitative tests of these mechanisms in non-magnetic CVs in the upcoming years are needed.

Turning this discussion around, do we expect to see jets in the magnetic CVs? The jet-framework outlined in Section~\ref{sec:motivationforjets} is fundamentally based on hysteresis in the accretion disk. If this scenario is accurate, then we do not expect jets in the most magnetic CVs that lack accretion discs (the polars). As discussed in \citet{Hameury2017}, the observed hysteresis in SS Cyg (Figure~\ref{fig:Hysteresis_diagram}; \citealt{Koerding2008}) is particularly a result of comparing the inner and outer accretion disk (or alternatively the inner accretion flow and the inner accretion disk, \citealt{Hameury2017}). The intermediate polars should not show this hysteresis, as they lack an inner accretion disc. Accordingly, we would also not expect intermediate polars to launch jets based on this jet-framework. For a model independent answer to whether magnetic CVs launch jets, VLBI observations of CVs are necessary.

%%%%%%%%%%%%%%%%%%%%%%%%%%%%%%%%%%%%%%%%%%%%%%%%%%%%%%%%%%%%%%%%%
\section{Future Work:}\label{sec:futurework}

To conclusively answer the question of whether CVs launch jets, there are a number of key observational and theoretical studies that need to be done. These include tests of the accretion-jet framework (in CVs and XRBs), model independent searches for jets, and thorough explorations of alternative radio emission mechanisms. Here we outline the work that (in our opinion) is the most important for progress in this field.

We start with observational studies of CVs. \textit{First}, multi-wavelength observations of a sample of DNe over the course of an outburst, with particular emphasis on the rise and decline phases, are necessary to test the accretion disk framework. At minimum, radio and optical observations will test the key prediction of the model: radio flaring during the state transition shortly after the rise to outburst. With EUV and X-ray observations to monitor the inner accretion flow, it will be possible to fully test the hysteresis in DNe - specifically whether all DNe outbursts progress through analogous accretion-outflow states to XRBs and whether a comparison of the emission through the inner accretion flow to the inner accretion disk (potentially a more accurate comparison to the XRBs, \citealt{Hameury2017}) produces the same hysteresis and associated outflows. To date, SS Cyg is the only CV with sufficient observations to test this. \textit{Second}, a distance-limited radio survey of CVs is needed to characterise the radio emission properties. Being able to define the radio luminosity, flaring time-scales, flaring amplitudes and polarization properties of a significant sample of CVs is critical for testing alternative radio emission mechanisms. \textit{Third}, high cadence radio observations of high inclination CVs over multiple orbits are needed to test if the radio emission is dependent on the orbital phase. This will (for example) indicate whether the emission is stemming from the secondary star, or whether it is larger scale emission. Current observations do not show this \citep{Coppejans2016b} but better sampling and longer observations are necessary, and there is evidence indicating a radio dependence on orbital period in some magnetic CVs \citep[e.g.][]{Mason2007,Gawronski2018}.

Looking further afield, complementary observations of other classes of objects are also necessary for context for the jet-accretion framework. Deep radio observations of NS XRBs in the soft-state will probe the degree of jet core quenching, and whether this is analogous to the soft-state radio emission seen in CVs. Currently sensitivity is a key limiting factor, but future generation telescopes such as the next generation Very Large Array (ngVLA) and Square Kilometre Array (SKA) will make a significant difference to this work. The focus up to now has been on a comparison of CVs to the XRBs, but a closer comparison to other white dwarf accretors is also warranted. Is the accretion-jet framework applicable to systems where there is an extra power source in the form of thermonuclear burning on the surface of the white dwarf or are we looking at different jet physics? Intriguingly, observations of the symbiotic star CH Cygni do show radio ejection events after the hard X-ray flux produced by the accretion flow drops \citep{Sokoloski2003}, which is analogous to what we see in SS Cyg and the XRBs \citep{Koerding2011}.

For a model independent answer to the question of whether CVs launch jets, a direct detection of a jet is necessary. This could be through VLBI observations or optical searches for satellite lines. Alternatively, will we see (and be able to distinguish) jet emission at other wavelengths, for example in the infrared? This has been suggested for the CV Z Cam, where \citet{Harrison2014} interpret the infrared emission to be synchrotron emission from a jet.

There are a number of facilities that have recently come online or are planned in the next decades that will help address these questions. The radio Karoo Array Telescope MeerKAT \citep{MeerKAT} and its linked optical telescope MeerLICHT \citep{MeerLICHT} will produce revolutionary simultaneous radio and optical light curves for every pointing. Furthermore, time on MeerKAT has already been approved for this CV science (through the ThunderKAT project, \citealt{Fender2017}). Looking beyond this, the SKA will have unparalleled sensitivity and resolution. The science case for the ngVLA \citep{ngVLA} is currently being reviewed -- its high-frequency coverage, sensitivity and resolution will drive key advances in this field \citep{Coppejans2018_ngVLA}. For complementary observations of the accretion flow (probed mainly by EUV, X-ray and optical wavelengths) telescopes such as Chandra, Evryscope, LSST, WFIRST, LUVOIR, Athena, STROBE-X, FINESSE and the Lynx X-ray Surveyor will be important.

On the theoretical side, there are also several key areas for investigation. The inner accretion flow and hysteresis in XRBs is not well understood (see Chapter 7 of this volume) and neither is the boundary layer in CVs. Yet a deep understanding of the inner accretion flow is necessary to accurately map the accretion-outflow hysteresis from one class to another (see Section \ref{sec:CV_HID}). Aside from the predictions of the accretion-jet framework, what other emission and/or timing properties can be used to identify (or rule out) jets in CVs? On the other hand, detailed models, tests and predictions for alternative radio emission mechanisms are also needed. For example, what would we expect a magnetic re-connection event in a rapidly rotating (period of a few hours), roche-lobe filling dwarf to look like? Is the local medium optically thin to such emission? Could this mechanism accelerate the electrons and produce the emission properties that we see in the non-magnetic CVs? If so, then can we explain the observed difference in the polarization properties between the non-magnetic and magnetic CVs? How do we expect the rate, amplitude and properties of potential magnetic reconnections in the accretion disk to change over the course of an outburst? Would we expect a high amplitude flare such as the one in SS Cyg to appear consistently at the start of every outburst, and can this mechanism explain the late high amplitude flares seen in some outbursts despite the comparatively empty state of the disk? Returning to CVs, what is the mechanism that produces anomalous outbursts and how do they relate to the DIM?

Determining whether CVs launch jets requires work and advances in a number of different fields. However, from the discussion here it is obvious that studies of CVs can provide a means to answer questions in related fields that are currently inaccessible. For example, their secondary stars may offer an environment to study stellar dynamo mechanisms at previously inaccessible spin rates. More directly relevant to this review, they offer an opportunity to test and constrain jet-launching physics in non-relativistic and comparatively well-understood environments with well-determined distances.    

%%%%%%%%%%%%%%%%%%%%%%%%%%%%%%%%%%%%%%%%%%%%%%%%%%%%%%%%%
\section{Conclusion}\label{sec:conclusion}

The question of whether CVs launch jets is not conclusively settled, yet the answer could have important implications for our understanding of jet physics. Indeed, a number of jet-launching models for accreting compact objects have been constrained assuming that CVs do not launch jets \citep[e.g.,][]{Livio1997,Livio1999,Soker2004}. Looking beyond this, they could be ideal targets for testing jet physics, as they are non-relativistic, show a range of physical properties (e.g. magnetic field strength) and have well-determined distances which enable an accurate determination of the jet properties. In this review we have outlined the theory and observations indicating that CVs launch jets. We then proceeded to discuss the open questions and issues with this framework that have been brought up by recent observations. Competing models for the radio emission were also discussed. In the last part of this review we highlighted key areas for future research towards conclusively answering whether CVs launch jets.

\section{Acknowledgements}

We thank the anonymous referee for their comments that have helped to improve this review. We would also like to thank James Miller-Jones, Jean-Pierre Lasota, Koji Mukai and Stuart Littlefair for useful discussions and suggestions.

%%%%%%%%%%%%%%%%%%%%%%%%%%%%%%%%%%%%%%%%%%%%%%%%%%%%%%%%%
\bibliographystyle{elsarticle-harv} 
\bibliography{master_bibliography}

\end{document}